# The Wilson-Racah Quantum System

A. D. Alhaidari[(1)†], T. J. Taiwo[(2)]

[(1)] *Saudi Center for Theoretical Physics, P.O. Box 32741, Jeddah 21438, Saudi Arabia*
[(2)] *Department of Mathematics, University of Benin, Benin City, Edo State 300283, Nigeria*

**Abstract**: Using a recent formulation of quantum mechanics without potential function, we present a four-parameter system associated with the Wilson and Racah polynomials. The continuum scattering states are written in terms of the Wilson polynomials whose asymptotics gives the scattering amplitude and phase shift. On the other hand, the finite number of discrete bound states are associated with the Racah polynomials.

We are honored to dedicate this work to Prof. Hashim A. Yamani on the occasion of his 70th birthday.

## 1. Introduction:

The connection between scattering and the asymptotics of orthogonal polynomials was described by Case and Geronimo [1-3]. Using these results, we introduced recently a formulation of quantum mechanics without the need for specifying a potential function [4]. The objective was to obtain a set of analytically realizable systems, which is larger than in the standard formulation and that may or may not be associated with any given or previously known potential functions. In this formulation, the wavefunction is written as an infinite bounded sum over a complete set of square integrable functions in configuration space. The expansion coefficients in the sum are orthogonal polynomials in the energy variable. Specifically, we write

$$\psi(E,x) = \sum_n f_n^\mu(E)\phi_n(x), \qquad (1)$$

where $\{\phi_n(x)\}_{n=0}^\infty$ is an $L^2$ basis functions in configuration space with coordinate $x$, $\{f_n^\mu(E)\}_{n=0}^\infty$ are the expansion coefficients at the energy $E$, and $\mu$ stands for a set of real parameters associated with the particular physical system. The basis set $\{\phi_n(x)\}_{n=0}^\infty$ satisfy the boundary conditions and contain only kinematical information (e.g., the angular momentum, a length scale, etc.). On the other hand, structural and dynamical information about the specific system under study is contained only in the expansion coefficients, which we write as $f_n^\mu(E) = f_0^\mu(E) P_n^\mu(\varepsilon)$, where $\varepsilon$ is some proper function of the energy. Thus, $P_0^\mu(\varepsilon) = 1$ and the completeness of the basis elements leads to the following orthogonality relation [4]

$$\int \rho^\mu(\varepsilon) P_n^\mu(\varepsilon) P_m^\mu(\varepsilon) d\varepsilon = \delta_{nm}, \qquad (2)$$

---

[†] Present Address: Al-Telal Project, Rawshan Villa #345, Al-Ranoona District, Madinah, Saudi Arabia

where $\rho^\mu(\varepsilon) = [f_0^\mu(E)]^2$ is the weight function in some appropriate energy domain. Thus, $\{P_n^\mu(\varepsilon)\}_{n=0}^\infty$ is a complete set of orthogonal polynomials in the energy. The physically relevant ones are those with the following asymptotics (limit as $n \to \infty$)

$$P_n^\mu(\varepsilon) \approx n^{-\tau} A(\varepsilon) \times \cos[n^\xi \theta(\varepsilon) + \delta(\varepsilon)], \tag{3}$$

where $\tau$ and $\xi$ are real positive constants that depend on the particular energy polynomial. The studies in [1-4] show that $A(\varepsilon)$ is the scattering amplitude and $\delta(\varepsilon)$ is the phase shift. Both depend not just on the energy but also on the set of physical parameters $\{\mu\}$. Bound states, if they exist, occur at energies $\{E_m\}$ that make the scattering amplitude vanish, $A(\varepsilon_m) = 0$. The number of these bound states is either finite or infinite and we write the $m^{\text{th}}$ bound state as

$$\psi(E_m, x) = \sqrt{\omega^\mu(\varepsilon_m)} \sum_n Q_n^\mu(\varepsilon_m) \phi_n(x), \tag{4}$$

where $\{Q_n^\mu(\varepsilon_m)\}$ are the discrete version of the polynomials $\{P_n^\mu(\varepsilon)\}$ and $\omega^\mu(\varepsilon_m)$ is the associated discrete weight function. In the absence of a potential function, the physical properties of the system in this formulation are deduced from the features of the orthogonal polynomials $\{P_n^\mu(\varepsilon), Q_n^\mu(\varepsilon_m)\}$. Such features include, but not limited to, the shape of the weight function, nature of the generating function, distribution and density of the polynomial zeros, recursion relation, asymptotics, differential or difference equations, etc.

In Ref. [4], the authors studied two- and three-parameter systems corresponding to the Meixner-Pollaczek polynomial and the continuous dual Hahn polynomial, respectively. Special cases of these systems include, but not limited to, the Coulomb, oscillator and Morse problems. Most notably though, new systems that do not belong to the already known class of exactly solvable problems were also found. Their associated scattering phase shift and bound states energy spectra were obtained analytically. In the following section, we introduce a four-parameter system associated with the Wilson polynomial and its discrete version, the Racah polynomial. We obtain the phase shift for the continuum scattering states and the energy spectrum for the bound states.

**2. The Wilson-Racah system:**

The total wavefunction of the system is written as $\Psi(t, x) = e^{-iEt/\hbar} \psi(E, x)$ giving the associated Hamiltonian as $H\Psi = i\hbar \frac{\partial}{\partial t} \Psi = E\Psi$. However, since a potential function is not given (or unknown) then $H$ cannot be written in the conventional quantum mechanical representation as the sum of the kinetic energy operator and a potential function. Nonetheless, all postulates of quantum mechanics still hold and all physical information about the system are contained in the wavefunction once a basis set $\{\phi_n(x)\}$ is chosen and the energy polynomials $\{P_n^\mu(\varepsilon)\}$ are specified. The basis elements will be chosen later, but now we select the four-parameter Wilson polynomial whose normalized version is shown in the Appendix as formula (A6). The corresponding normalized weight function is given by (A5). Comparing the asymptotic formula (A9) with Eq. (3) and noting that $\ln n \approx o(n^\xi)$ for any $\xi > 0$, we conclude that the scattering phase shift is

$$\delta(\varepsilon) = \arg\{\Gamma(2iy)/\Gamma(\mu+iy)\Gamma(\nu+iy)\Gamma(a+iy)\Gamma(b+iy)\}, \tag{5}$$

where $y = \varepsilon(E)$ such that $y \geq 0$. If we choose $\varepsilon = k/\lambda$, where $E = \frac{1}{2}k^2$ and $\lambda^{-1}$ is the length scale of the system in the atomic units $\hbar = m = 1$, then $y = \varepsilon = \sqrt{2E}/\lambda$ and Fig. 1 is a plot of the scattering phase shift for a given choice of values of the physical parameters. If all parameters are positive, then there are no bound states. However, if $\mu < 0$ and $\mu + \nu$, $\mu + a$, $\mu + b$ are positive then in addition to the continuum scattering states there exist $N$ bound states, where $N$ is the largest non-negative integer less than or equal to $-\mu$. These bound states occur at energies $\{\varepsilon_m\}_{m=0}^{N}$ such that $iy = -(m+\mu)$ which makes the scattering amplitude $A(\varepsilon)$ vanish since the argument of the gamma function $\Gamma(\mu+iy)$ in the denominator of $|\mathcal{A}(iy)|$ in Eq. (A9) becomes a non-positive integer. Consequently, the bound states energies are obtained from the energy spectrum formula $y^2 = -(m+\mu)^2$ giving

$$E_m = -\frac{\lambda^2}{2}(m+\mu)^2, \tag{6}$$

and the continuous orthogonality relation (A4) will be augmented by a discrete part and becomes [5]

$$\frac{1}{2\pi}\frac{\Gamma(\mu+\nu+a+b)}{\Gamma(\mu+\nu)\Gamma(a+b)\Gamma(\mu+a)\Gamma(\mu+b)\Gamma(\nu+a)\Gamma(\nu+b)} \times$$
$$\int_0^\infty |\Gamma(\mu+iy)\Gamma(\nu+iy)\Gamma(a+iy)\Gamma(b+iy)/\Gamma(2iy)|^2 \, W_n^\mu(y^2;\nu,a,b)\,W_{n'}^\mu(y^2;\nu,a,b)\,dy$$
$$-2\frac{\Gamma(\mu+\nu+a+b)\Gamma(\nu-\mu)\Gamma(a-\mu)\Gamma(b-\mu)}{\Gamma(-2\mu+1)\Gamma(a+b)\Gamma(a+\nu)\Gamma(b+\nu)} \times$$
$$\sum_{m=0}^{N}(m+\mu)\frac{(2\mu)_m(\mu+\nu)_m(\mu+a)_m(\mu+b)_m}{(\mu-\nu+1)_m(\mu-a+1)_m(\mu-b+1)_m m!} W_n^\mu(-(m+\mu)^2;\nu,a,b)\,W_{n'}^\mu(-(m+\mu)^2;\nu,a,b) = \delta_{n,n'} \tag{7}$$

The total wavefunction corresponding to the continuous energy $\varepsilon$ and discrete energy $\varepsilon_m$ is written as

$$\psi_m(E,x) = \sqrt{\rho^\mu(\varepsilon)} \sum_{n=0}^{\infty} W_n^\mu(\varepsilon^2;\nu,a,b)\phi_n(x)$$
$$+ \sqrt{\rho_m^N} \sum_{n=0}^{N} W_n^\mu(-(m+\mu)^2;\nu,a,b)\phi_n(x) \tag{8}$$

where $\rho^\mu(\varepsilon)$ and $\rho_m^N$ are the continuous and discrete normalized weight functions deduced from the orthogonality relation (7).

On the other hand, there are other possibilities that make $A(\varepsilon) = 0$ such as $\nu + iy = -m$, $a + iy = -m$ and/or $b + iy = -m$. The corresponding energy spectrum formula will be $y^2 = -(m+\nu)^2$, $y^2 = -(m+a)^2$ and/or $y^2 = -(m+b)^2$. Now, if the values of the physical parameters are such that the system is totally confined (for example, $\mu < 0$, $1 > \nu \geq 0$, $\mu + a > 0$, and $\mu + b > 0$), then there are only bound states and the $m^{\text{th}}$ wavefunction will be written in terms of the discrete Racah polynomial [6] as follows

$$\psi_m(x) = \sqrt{\rho^N(m;\alpha,\beta,\gamma)} \sum_{n=0}^{N} R_n^N(m;\alpha,\beta,\gamma)\phi_n(x), \tag{9}$$

where $\rho^N(m;\alpha,\beta,\gamma)$ and $R_n^N(m;\alpha,\beta,\gamma)$ are given in the Appendix by Eq. (A15) and Eq. (A16), respectively. The totally discrete orthogonality relation is (A17). Figure 2 is a plot of the lowest bound states for a given set of physical parameters $\{\alpha,\beta,\gamma\}$ and where the basis elements are chosen in the one-dimensional configuration space with coordinate $-\infty < x < +\infty$ as follows

$$\phi_n(x) = [\sqrt{\pi} 2^n n!]^{-1/2} e^{-\lambda^2 x^2/2} H_n(\lambda x), \qquad (10)$$

where $H_n(z)$ is the Hermite polynomial of degree $n$ in $z$. Figure 3 is for the same system but in three dimensions with spherical symmetry and radial coordinate $r$ and where the basis elements are chosen as

$$\phi_n(r) = \sqrt{\frac{\Gamma(n+1)}{\Gamma(n+2\ell+2)}} (\lambda r)^{\ell+\frac{1}{2}} e^{-\lambda r/2} L_n^{2\ell+1}(\lambda r), \qquad (11)$$

with $\ell$ being the angular momentum quantum number and $L_n^\nu(z)$ is the associated Laguerre polynomial of degree $n$ in $z$.

## 3. Conclusion and discussion:

Using the formulation of quantum mechanics without a potential function, we introduced here a four-parameter system whose continuum scattering states are associated with the Wilson polynomial and the discrete bound states are associated with the Racah polynomial. These polynomials constitute the expansion coefficients of the wavefunction in a complete set of square integrable basis elements in configuration space. Depending on the values of the physical parameters, the system consist of scattering states and/or a finite number of bound states. The scattering phase shift and energy spectrum were obtained analytically.

Finally, we note that the choice of the polynomial argument $y$ as a function of the energy is not unique. Making another choice will result in a physically different system with different scattering phase shift and energy spectrum. For example, choosing $y = \lambda/k$ gives the energy spectrum formula $E_m = -\lambda^2/2(m+\mu)^2$. Whereas, if we take $y = \sqrt{\ln(1+k^2/\lambda^2)}$ then we obtain the following energy spectrum

$$E_m = \frac{\lambda^2}{2}[e^{-(m+\mu)^2} - 1]. \qquad (12)$$

**Acknowledgements**:

We like to acknowledge the support by the Saudi Center for Theoretical Physics (SCTP) during the progress of this work.

**Appendix: The Wilson and Racah polynomials**

The Wilson polynomial, $\tilde{W}_n^\mu(y^2;\nu,a,b)$, is defined here as [5]

$$\tilde{W}_n^\mu(y^2;\nu,a,b) = \frac{(\mu+a)_n(\mu+b)_n}{(a+b)_n n!} {}_4F_3\left(\begin{matrix}-n, n+\mu+\nu+a+b-1, \mu+iy, \mu-iy \\ \mu+\nu, \mu+a, \mu+b\end{matrix}\bigg|1\right), \qquad (A1)$$

where $_4F_3\left(\begin{array}{c}a,b,c,d\\e,f,g\end{array}\bigg|z\right) = \sum_{n=0}^{\infty} \frac{(a)_n(b)_n(c)_n(d)_n}{(e)_n(f)_n(g)_n n!} z^n$ is the hypergeometric function and $(a)_n = a(a+1)(a+2)...(a+n-1) = \frac{\Gamma(n+a)}{\Gamma(a)}$. The generating function of these polynomials is

$$\sum_{n=0}^{\infty} \tilde{W}_n^\mu(y^2;v;a,b) t^n = {}_2F_1\left(\begin{array}{c}\mu+iy,v+iy\\\mu+v\end{array}\bigg|t\right) {}_2F_1\left(\begin{array}{c}a-iy,b-iy\\a+b\end{array}\bigg|t\right), \quad (A2)$$

and their three-term recursion relation (for $n=1,2,3,...$) is

$$y^2 \tilde{W}_n^\mu = \left[\frac{(n+\mu+v)(n+\mu+a)(n+\mu+b)(n+\mu+v+a+b-1)}{(2n+\mu+v+a+b)(2n+\mu+v+a+b-1)} + \frac{n(n+v+a-1)(n+v+b-1)(n+a+b-1)}{(2n+\mu+v+a+b-1)(2n+\mu+v+a+b-2)} - \mu^2\right] \tilde{W}_n^\mu \quad (A3)$$
$$- \frac{(n+\mu+a-1)(n+\mu+b-1)(n+v+a-1)(n+v+b-1)}{(2n+\mu+v+a+b-1)(2n+\mu+v+a+b-2)} \tilde{W}_{n-1}^\mu - \frac{(n+1)(n+\mu+v)(n+a+b)(n+\mu+v+a+b-1)}{(2n+\mu+v+a+b)(2n+\mu+v+a+b-1)} \tilde{W}_{n+1}^\mu$$

The initial seeds ($n=0$) for this recursion are $\tilde{W}_0^\mu = 1$ and $\tilde{W}_1^\mu = \frac{(\mu+a)(\mu+b)}{a+b} - \frac{\mu+v+a+b}{(\mu+v)(a+b)} (y^2+\mu^2)$. If $\mathrm{Re}(\mu,v,a,b) > 0$ and non-real parameters occur in conjugate pairs, then the orthogonality relation becomes

$$\frac{1}{2\pi}\int_0^\infty \frac{\Gamma(\mu+v+a+b)|\Gamma(\mu+iy)\Gamma(v+iy)\Gamma(a+iy)\Gamma(b+iy)|^2}{\Gamma(\mu+v)\Gamma(a+b)\Gamma(\mu+a)\Gamma(\mu+b)\Gamma(v+a)\Gamma(v+b)|\Gamma(2iy)|^2} \tilde{W}_n^\mu(y^2;v;a,b)\tilde{W}_m^\mu(y^2;v;a,b) dy \quad (A4)$$
$$= \left(\frac{n+\mu+v+a+b-1}{2n+\mu+v+a+b-1}\right) \frac{(\mu+a)_n(\mu+b)_n(v+a)_n(v+b)_n}{(\mu+v)_n(a+b)_n(\mu+v+a+b)_n n!} \delta_{nm}$$

Thus, the normalized weight function is

$$\rho^\mu(y;v;a,b) = \frac{1}{2\pi} \frac{\Gamma(\mu+v+a+b)|\Gamma(\mu+iy)\Gamma(v+iy)\Gamma(a+iy)\Gamma(b+iy)/\Gamma(2iy)|^2}{\Gamma(\mu+v)\Gamma(a+b)\Gamma(\mu+a)\Gamma(\mu+b)\Gamma(v+a)\Gamma(v+b)}. \quad (A5)$$

and the orthonormal version of the polynomial is

$$W_n^\mu(y^2;v;a,b) = \sqrt{\left(\frac{2n+\mu+v+a+b-1}{n+\mu+v+a+b-1}\right)\frac{(\mu+v)_n(a+b)_n(\mu+v+a+b)_n n!}{(\mu+a)_n(\mu+b)_n(v+a)_n(v+b)_n}} \tilde{W}_n^\mu(y^2;v;a,b)$$
$$= \sqrt{\left(\frac{2n+\mu+v+a+b-1}{n+\mu+v+a+b-1}\right)\frac{(\mu+a)_n(\mu+b)_n(\mu+v)_n(\mu+v+a+b)_n}{(v+a)_n(v+b)_n(a+b)_n n!}} \, {}_4F_3\left(\begin{array}{c}-n,n+\mu+v+a+b-1,\mu+iy,\mu-iy\\\mu+v,\mu+a,\mu+b\end{array}\bigg|1\right) \quad (A6)$$

The three-term recursion relation for this orthonormal version of the polynomial is symmetric and reads as follows:

$$y^2 W_n^\mu = \left[\frac{(n+\mu+v)(n+\mu+a)(n+\mu+b)(n+\mu+v+a+b-1)}{(2n+\mu+v+a+b)(2n+\mu+v+a+b-1)} + \frac{n(n+v+a-1)(n+v+b-1)(n+a+b-1)}{(2n+\mu+v+a+b-1)(2n+\mu+v+a+b-2)} - \mu^2\right] W_n^\mu$$
$$- \frac{1}{2n+\mu+v+a+b-2}\sqrt{\frac{n(n+\mu+v-1)(n+a+b-1)(n+\mu+a-1)(n+\mu+b-1)(n+v+a-1)(n+v+b-1)(n+\mu+v+a+b-2)}{(2n+\mu+v+a+b-3)(2n+\mu+v+a+b-1)}} W_{n-1}^\mu \quad (A7)$$
$$- \frac{1}{2n+\mu+v+a+b}\sqrt{\frac{(n+1)(n+\mu+v)(n+a+b)(n+\mu+a)(n+\mu+b)(n+v+a)(n+v+b)(n+\mu+v+a+b-1)}{(2n+\mu+v+a+b-1)(2n+\mu+v+a+b+1)}} W_{n+1}^\mu$$

Using the results of Wilson's work [7] and those in [8], we obtain the following asymptotics ($n \to \infty$)

$$\tilde{W}_n^\mu(y^2;v;a,b) \approx \frac{1}{n}\Gamma(\mu+v)\Gamma(a+b)\left\{2|\mathcal{A}(iy)|\cos[2y\ln n + \arg\mathcal{A}(iy)] + O(n^{-1})\right\}. \quad (A8)$$

Or equivalently,

$$W_n^\mu(y^2;v;a,b) \approx B(\mu,v,a,b)\sqrt{\frac{2}{n}}\left\{2|\mathcal{A}(iy)|\cos[2y\ln n + \arg\mathcal{A}(iy)] + O(n^{-1})\right\}, \quad (A9)$$

where $\mathcal{A}(z) = \Gamma(2z)/\Gamma(\mu+z)\Gamma(v+z)\Gamma(a+z)\Gamma(b+z)$ and $B(\mu,v,a,b) = \sqrt{\Gamma(\mu+v)\Gamma(a+b)\Gamma(\mu+a)\Gamma(\mu+b)\Gamma(v+a)\Gamma(v+b)/\Gamma(\mu+v+a+b)}$. Therefore, discrete bound states exist if the scattering amplitude vanishes, which occurs, for example, if $\mu + iy = -m$ and $m = 0,1,2,..,N$. Analysis of bound states show that we should have $\mu < 0$ with $N$ being the largest non-negative integer less than or equal to $-\mu$. Then $\tilde{W}_n^\mu(y^2; v; a, b)$ becomes the Racah polynomial, $\tilde{R}_n^N(m; \alpha, \beta, \gamma)$, which is defined here as [6]

$$\tilde{R}_n^N(m; \alpha, \beta, \gamma) = \frac{(\alpha+1)_n (\gamma+1)_n}{(\alpha+\beta+N+2)_n n!} {}_4F_3\left(\begin{array}{c}-n,-m,n+\alpha+\beta+1,m-\beta+\gamma-N\\ \alpha+1,\gamma+1,-N\end{array}\Big|1\right). \tag{A10}$$

The parameter map that takes (A1) into (A10) is: $\alpha = \mu + a - 1$, $\gamma = \mu + b - 1$, and $\beta = v + b - 1$. The extra parameter $\delta$ in [6] is related to $N$ as $\delta = -(N + \beta + 1) = \mu - b$ and the parameter constraints are: $\alpha > -1$, $\gamma > -1$, $\beta > N - 1$. The inverse parameter map is: $\mu = \frac{1}{2}(\gamma + \delta + 1)$, $v = \beta + \frac{1}{2}(\delta - \gamma + 1)$, $a = \alpha - \frac{1}{2}(\gamma + \delta - 1)$ and $b = \frac{1}{2}(\gamma - \delta + 1)$. The generating function of these polynomials is obtained from (A2) using the parameter map giving

$$\sum_{n=0}^N \tilde{R}_n^N(m; \alpha, \beta, \gamma) t^n = {}_2F_1\left(\begin{array}{c}-m,-m+\beta-\gamma\\-N\end{array}\Big|t\right) {}_2F_1\left(\begin{array}{c}m+\alpha+1,m+\gamma+1\\ \alpha+\beta+N+2\end{array}\Big|t\right), \tag{A11}$$

which is formula (1.2.12) in [6]. The three-term recursion relation is

$$\tfrac{1}{4}(N+\beta-\gamma-2m)^2 \tilde{R}_n^N =$$

$$\left[\tfrac{1}{4}(N+\beta-\gamma)^2 - \frac{(n-N)(n+\alpha+1)(n+\gamma+1)(n+\alpha+\beta+1)}{(2n+\alpha+\beta+1)(2n+\alpha+\beta+2)} - \frac{n(n+\beta)(n+\alpha+\beta-\gamma)(n+N+\alpha+\beta+1)}{(2n+\alpha+\beta)(2n+\alpha+\beta+1)}\right]\tilde{R}_n^N \tag{A12}$$

$$+ \frac{(n+\alpha)(n+\beta)(n+\gamma)(n+\alpha+\beta-\gamma)}{(2n+\alpha+\beta)(2n+\alpha+\beta+1)} \tilde{R}_{n-1}^N + \frac{(n+1)(n-N)(n+\alpha+\beta+1)(n+N+\alpha+\beta+2)}{(2n+\alpha+\beta+1)(2n+\alpha+\beta+2)} \tilde{R}_{n+1}^N$$

which is equivalent to formula (1.2.3) in [6]. The discrete orthogonality reads as follows

$$\sum_{m=0}^N \frac{2m+\gamma-\beta-N}{m+\gamma-\beta-N} \frac{(-N)_m (\alpha+1)_m (\gamma+1)_m (\gamma-\beta-N+1)_m}{(-\beta-N)_m (\gamma-\beta+1)_m (\gamma-\alpha-\beta-N)_m m!} \bar{R}_n^N(m;\alpha,\beta,\gamma) \bar{R}_{n'}^N(m;\alpha,\beta,\gamma)$$

$$= \frac{n+\alpha+\beta+1}{2n+\alpha+\beta+1} \frac{(-\alpha-\beta-N-1)_N (\gamma-\beta-N+1)_N}{(-\beta-N)_N (\gamma-\alpha-\beta-N)_N} \frac{(\beta+1)_n (\alpha+\beta-\gamma+1)_n (\alpha+\beta+N+2)_n n!}{(-N)_n (\alpha+1)_n (\gamma+1)_n (\alpha+\beta+2)_n} \delta_{n,n'} \tag{A13}$$

which is formula (1.2.2) in [6] and where

$$\bar{R}_n^N(m;\alpha,\beta,\gamma) = {}_4F_3\left(\begin{array}{c}-n,-m,n+\alpha+\beta+1,m-\beta+\gamma-N\\ \alpha+1,\gamma+1,-N\end{array}\Big|1\right). \tag{A14}$$

Thus, the discrete normalized weight function is

$$\rho^N(m;\alpha,\beta,\gamma) = \frac{(-\beta-N)_N (\gamma-\alpha-\beta-N)_N}{(-\alpha-\beta-N-1)_N (\gamma-\beta-N+1)_N} \times$$

$$\frac{2m+\gamma-\beta-N}{m+\gamma-\beta-N} \frac{(-N)_m (\alpha+1)_m (\gamma+1)_m (\gamma-\beta-N+1)_m}{(-\beta-N)_m (\gamma-\beta+1)_m (\gamma-\alpha-\beta-N)_m m!} \tag{A15}$$

and the orthonormal version of the discrete Racah polynomial is

$$R_n^N(m;\alpha,\beta,\gamma) = \sqrt{\frac{2n+\alpha+\beta+1}{n+\alpha+\beta+1} \frac{(-N)_n (\alpha+1)_n (\gamma+1)_n (\alpha+\beta+2)_n}{(\beta+1)_n (\alpha+\beta-\gamma+1)_n (\alpha+\beta+N+2)_n n!}} \tag{A16}$$

$${}_4F_3\left(\begin{array}{c}-n,-m,n+\alpha+\beta+1,m-\beta+\gamma-N\\ \alpha+1,\gamma+1,-N\end{array}\Big|1\right)$$

Thus, we can rewrite (A13) as

$$\sum_{m=0}^N \rho^N(m;\alpha,\beta,\gamma) R_n^N(m;\alpha,\beta,\gamma) R_{n'}^N(m;\alpha,\beta,\gamma) = \delta_{n,n'}. \tag{A17}$$

In these calculations, we used the identities $\frac{(a+1)_n}{(a)_n} = \frac{n+a}{a}$, $\frac{(a)_{n+1}}{(a)_n} = n+a$, $(n+a)_n = \frac{\Gamma(2n+a)}{\Gamma(n+a)}$, $\frac{(n+a)_n}{(a+1)_{2n}} = \frac{a/(a)_n}{2n+a}$, and $(a-n)_n = \frac{\Gamma(a)}{\Gamma(a-n)}$. An orthogonality relation, which is dual to (A13) is obtained by utilizing the interchange symmetry ($n \leftrightarrow m$, $\alpha \leftrightarrow \gamma$ and $\beta \leftrightarrow \delta$) in the definition of the polynomial. Performing this interchange in the original orthogonality relation (A13) gives its dual as follows

$$\sum_{m=0}^{N} \frac{2m+\alpha+\beta+1}{m+\alpha+\beta+1} \frac{(-N)_m (\alpha+1)_m (\gamma+1)_m (\alpha+\beta+2)_m}{(\beta+1)_m (\alpha+\beta-\gamma+1)_m (\alpha+\beta+N+2)_m m!} \bar{R}_m^N(n;\alpha,\beta,\gamma) \bar{R}_m^N(n';\alpha,\beta,\gamma)$$
$$= \frac{n+\gamma-\beta-N}{2n+\gamma-\beta-N} \frac{(\alpha+\beta+2)_N (\beta-\gamma)_N}{(\beta+1)_N (\alpha+\beta-\gamma+1)_N} \frac{(-\beta-N)_n (\gamma-\beta+1)_n (\gamma-\alpha-\beta-N)_n n!}{(-N)_n (\alpha+1)_n (\gamma+1)_n (\gamma-\beta-N+1)_n} \delta_{n,n'} \quad \text{(A18)}$$

Therefore, if we write the original orthogonality (A13) as

$$\sum_{m=0}^{N} \omega^N(m) \bar{R}_n^N(m;\alpha,\beta,\gamma) \bar{R}_{n'}^N(m;\alpha,\beta,\gamma) = \lambda^N \frac{\delta_{n,n'}}{\hat{\omega}^N(n)}. \quad \text{(A19a)}$$

Then, the dual orthogonality (A18) could be rewritten as follows

$$\sum_{m=0}^{N} \hat{\omega}^N(m) \bar{R}_m^N(n;\alpha,\beta,\gamma) \bar{R}_m^N(n';\alpha,\beta,\gamma) = \hat{\lambda}^N \frac{\delta_{n,n'}}{\omega^N(n)}, \quad \text{(A19b)}$$

where

$$\omega^N(m) = \frac{2m+\gamma-\beta-N}{m+\gamma-\beta-N} \frac{(-N)_m (\alpha+1)_m (\gamma+1)_m (\gamma-\beta-N+1)_m}{(-\beta-N)_m (\gamma-\beta+1)_m (\gamma-\alpha-\beta-N)_m m!}, \quad \text{(A20a)}$$

$$\hat{\omega}^N(m) = \frac{2m+\alpha+\beta+1}{m+\alpha+\beta+1} \frac{(-N)_m (\alpha+1)_m (\gamma+1)_m (\alpha+\beta+2)_m}{(\beta+1)_m (\alpha+\beta-\gamma+1)_m (\alpha+\beta+N+2)_m m!}, \quad \text{(A20b)}$$

$$\lambda^N = \frac{(-\alpha+\delta)_N (\gamma+\delta+2)_N}{(\delta+1)_N (\gamma-\alpha+\delta+1)_N} = \frac{(-\alpha-\beta-N-1)_N (\gamma-\beta-N+1)_N}{(-\beta-N)_N (\gamma-\alpha-\beta-N)_N}. \quad \text{(A21)}$$

Moreover, $\hat{\omega}^N(m) = \omega^N(m)\big|_{\substack{\alpha \leftrightarrow \gamma \\ \beta \leftrightarrow \delta}}$ and $\hat{\lambda}^N = \lambda^N\big|_{\substack{\alpha \leftrightarrow \gamma \\ \beta \leftrightarrow \delta}}$.

**Figures Caption**

**Fig. 1**: The scattering phase shift (in units of $\pi$) for the Wilson quantum system with physical parameters: $\mu = 0.7$, $\nu = 0.2$, $a = 0.5$, and $b = 0.3$. The energy variable is $\varepsilon = k/\lambda$, where $E = \frac{1}{2}k^2$ and $\lambda^{-1}$ is the system's length scale.

**Fig. 2**: The lowest bound states for the Racah quantum system with physical parameters: $N = 10$, $\alpha = 0.7$, $\gamma = 0.5$, and $\beta = N + 0.3$. The black, red, green, and brown trace corresponds to $m = 0, 1, 2,$ and 3, respectively. The basis elements in configuration space are given by Eq. (10) and the coordinate $x$ is measured in units of the length scale $\lambda^{-1}$.

**Fig. 3**: The same physical system as in Fig. 2 but in three dimensions with spherical symmetry and radial coordinate $r$ which is also measured in units of $\lambda^{-1}$. The basis elements are given by Eq. (11) and we took the angular momentum quantum number $\ell = 1$.

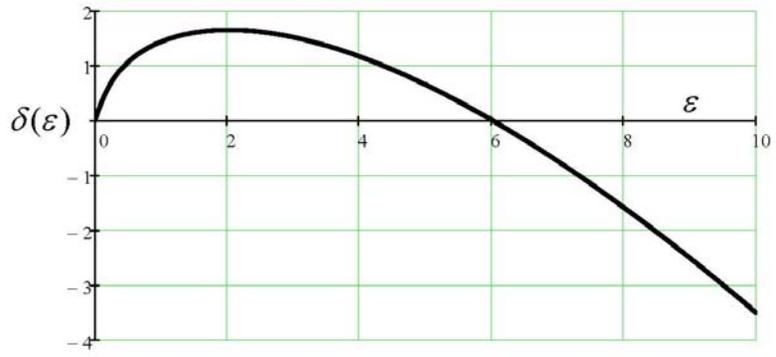

**Fig. 1**

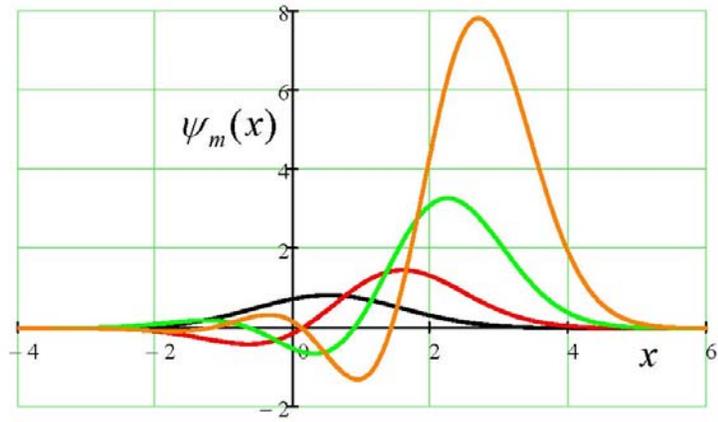

**Fig. 2**

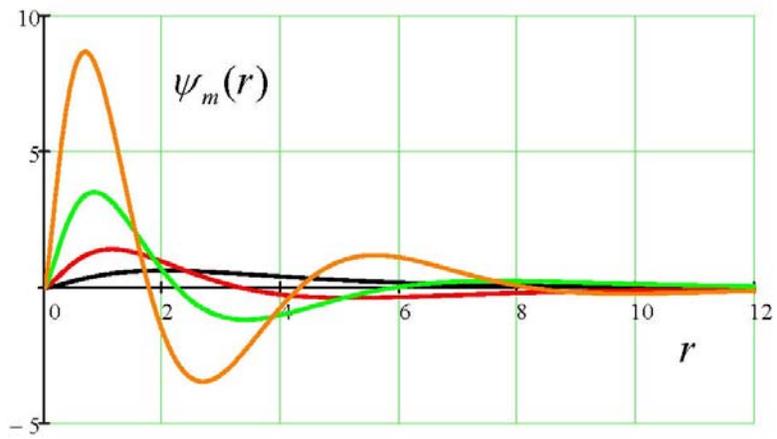

**Fig. 3**